\documentclass[conference]{IEEEtran}

\usepackage{hyperref}
\hypersetup{
     colorlinks = true,
     citecolor = green,
     }

\ifCLASSINFOpdf
  \usepackage[pdftex]{graphicx}
\else
\fi

\usepackage{framed}
\usepackage{subcaption}

\usepackage{xcolor}
\definecolor{cs-blue}{rgb}{0.035, 0.114, 0.235}
\definecolor{turq}{rgb}{0.365, 0.835, 0.851}

\usepackage{tcolorbox}

\tcbset{
  takeawaybox/.style={
    colback=turq!7!white, 
    colframe=turq, 
    coltitle=black, 
    fonttitle=\bfseries\small, 
    title=Key Takeaways: 
  }
}

\usepackage{balance}
\balance

\begin{document}

\title{Trust Calibration in IDEs: Paving the Way for Widespread Adoption of AI Refactoring}

\author{
\IEEEauthorblockN{Markus Borg}
\IEEEauthorblockA{\textit{CodeScene and Lund University}\\
Malmö, Sweden \\
markus.borg@codescene.com}
}

%


\maketitle

\begin{abstract}
In the software industry, the drive to add new features often overshadows the need to improve existing code. Large Language Models (LLMs) offer a new approach to improving codebases at an unprecedented scale through AI-assisted refactoring. However, LLMs come with inherent risks such as braking changes and the introduction of security vulnerabilities. We advocate for encapsulating the interaction with the models in IDEs and validating refactoring attempts using trustworthy safeguards. However, equally important for the uptake of AI refactoring is research on trust development. In this position paper, we position our future work based on established models from research on human factors in automation. We outline action research within CodeScene on development of 1) novel LLM safeguards and 2) user interaction that conveys an appropriate level of trust. The industry collaboration enables large-scale repository analysis and A/B testing to continuously guide the design of our research interventions.
\end{abstract}


%

\section{Introduction} \label{sec:intro}
Large Language Models (LLMs) have quickly become an integral part of Software Engineering (SE) practice. JetBrain's annual developer survey 2023 revealed that already a year ago, 77\% of developers use ChatGPT and 46\% use GitHub Copilot~\cite{jetbrains_state_2023}. While AI promises a coding revolution with tools to expedite code writing, we posit that the core challenge lies elsewhere. Developers spend most of their time understanding and maintaining existing code~\cite{minelli_i_2015}. While high-quality code is known to improve productivity~\cite{tornhill_code_2022}, reduce defect counts and vulnerabilities~\cite{al-boghdady_presence_2021}, and increase developer satisfaction~\cite{graziotin_unhappiness_2017}, organizations routinely prioritize implementing new features over improving existing code~\cite{li_systematic_2015}. Our premise is that there is substantial potential in applying LLMs to improve the maintainability of existing code -- we call this \textit{AI refactoring}.

Despite the potential of LLMs, they come with inherent risks. Research shows that LLMs’ prompt-based generation of non-deterministic code poses several dangers~\cite{yetistiren_assessing_2022,asare_is_2023}. Within the context of AI refactoring, these risks manifest as potential security vulnerabilities and unexpected changes in functional behavior. Our own investigation into four function-level JavaScript code smells exposed that the raw output from state-of-the-art LLMs provides correct refactorings in only 37\% of cases~\cite{tornhill_refactoring_2024}, a finding corroborated in recent work by Pomian \textit{et al.}~\cite{pomian_next-generation_2024}. Furthermore, we have observed several instances where subtle defects were introduced. We advocate turning the large numbers of blunt refactoring recommendations into a reduced set of sharp edits, effectively prioritizing precision over recall in line with recommendations from Google Research~\cite{froemmgen_resolving_2024}.

To address the unreliable nature of raw LLM output, we advocate for encapsulating the interaction with the models in IDEs. We discourage developers from refactoring code by copying solutions from LLM interactions in web browsers. Instead, AI refactoring should seamlessly integrate into IDEs -- as pioneered by GitHub Copilot -- and come equipped with \textit{safeguards} to validate their output. We draw inspiration from our previous work on using safety cages to enable the safe use of deep learning in automotive systems~\cite{borg_ergo_2023}. In context of AI refactoring, we argue that research on effective \textit{safeguards} is a necessary step toward more capable AI refactoring that transcends the function level.

AI refactoring within IDEs presents an ideal point for targeted research interventions. The IDE can act as the first gatekeeper for LLM output before it gets committed to the codebase. Additionally, studying IDE-level interventions allows for fine-grained data collection, guiding UI design for an optimal developer experience -- crucial for acceptance and realizing our vision of widespread adoption of AI refactoring.

However, we posit that \textit{trustworthiness} is only one side of the coin. Equally important for the uptake of AI refactoring is user \textit{trust}. Analogous to previous paradigm shifts, e.g., mechanization and electrification, AI uptake requires users to trust the technology. In the AI refactoring context, developers must develop trust in the technology to accept increasing automation levels. Developers are a diverse population, and the trust development journey can vary widely across individuals, influenced by their experiences, preferences, and work contexts.

We argue that the IDE is the best arena to foster developer trust. As the developers' primary workspace, it provides a familiar and effective starting point. However, building trust is a complex process, and research is needed to enable trust calibration within the IDE -- achieving the right balance between trust and trustworthiness. An imbalance can result in either reckless reliance on AI refactoring or unnecessary delays in adopting a highly valuable tool. Addressing this challenge demands interdisciplinary research that bridges human factors in automation, human-machine interaction, and computer science.

The rest of the paper is organized as follows. Section~\ref{sec:rw} presents related work on LLM-based software maintenance and recommendation delivery. Next, we present how we build on the theoretical background related to trust and its development in Sec.~\ref{sec:trust} and~\ref{sec:trustdev}, respectively. Finally, Sec.~\ref{sec:conc} presents a short conclusion and outlines our next steps.

\section{Related Work} \label{sec:rw}
This section presents an overview of research on LLM-based software maintenance and recommendation delivery in IDEs. Amid an avalanche of academic papers on LLM and SE, two sets of researchers have undertaken ambitious literature reviews to assess the current landscape. Hou \textit{et al.} identified 229 papers in a systematic literature review~\cite{hou_large_2024} and Fan \textit{et al.} found 244 related preprints on arXiv~\cite{fan_large_2023}. Both papers list primary studies that use LLMs for various SE tasks, including code generation, testing, maintenance, and documentation. Within maintenance, Automatic Program Repair (APR) emerges as the most common application, while refactoring has received less attention.

We investigated the subset of refactoring and APR papers to explore 1) how proposed patches were validated for correctness and 2) which datasets were studied. Our findings reveal that most papers rely on successful compilation and passing unit tests for validation, despite research highlighting concerns regarding the effectiveness of test suites~\cite{vercammen_mutation_2024}. In terms of static code analysis, we found examples of vulnerability analysis~\cite{pearce_examining_2023}, abstract syntax tree checks~\cite{tufano_empirical_2019}, and similarity measurements~\cite{fakhoury_towards_2023}. Finally, most papers target small Java and Python datasets, i.e., there is a need for larger studies.

In the context of refactoring assistance, there has been a notable evolution from \textit{recommendation systems} (RecSys)~\cite{bavota_recommending_2014,wyrich_towards_2019} to \textit{software bots}~\cite{santhanam_bots_2022}. While both approaches offer developers guidance, bots typically provide higher levels of automation~\cite{parasuraman_model_2000}, often encompassing task execution, and frequently engage in dynamic two-way interactions. Regardless of the approach, an effective UI is essential for conveying recommendations and rationales. Google's latest work on their internal reviewing assistant stresses how they successfully balanced moderate LLM results with a carefully crafted UI~\cite{froemmgen_resolving_2024}.

A critical UI consideration is the delivery of patch proposals to developers. UI design remains a very active topic in the general RecSys community, with calls for more practitioner-led research to investigate underlying questions such as how ``various implementations affect design qualities such as trust, fun, transparency, serendipity, sense of control''~\cite{smits_towards_2023}. In the RecSys SE literature, fundamental guidelines include designing for understandability, transparency, and assessability before considering trust cultivation~\cite{murphy-hill_recommendation_2014}. With the increasing focus on bots, interaction has taken the front seat. Non-intrusiveness was a topic already for RecSys, but this turns even more important in the bot setting~\cite{brown_sorry_2019}. In the same vein, it is important not to overload the developers with noise~\cite{wessel_dont_2021}. To conclude, while existing guidelines for designing APR bots will support our work on AI refactoring~\cite{smits_towards_2023}, the literature clearly motivates further research on UI design for developer acceptance and trust~\cite{wyrich_perception_2020}.

\section{Trust in the Context of AI Refactoring} \label{sec:trust}
Various research fields have studied the socio-psychological concept of \textit{trust}, and many researchers have tried explaining the multi-faceted concept. A systematic review by Hoff and Bashir reports that most explanations contain three constituents~\cite{hoff_trust_2015}. First, there must be a \textit{truster} to give trust (e.g., AI assistant), a \textit{trustee} to accept the trust (e.g., a developer), and something must be \textit{at stake} (e.g., code integrity). Second, the trustee must have an \textit{incentive} to perform some task (e.g., code smell removal). Third, there must be a \textit{risk} that the trustee will fail to perform the task, leading to potential consequences (e.g., breaking the build). 

Trust is related to \textit{reliance}, explained by Baier as ``continued relationship on the basis of dependable habits''~\cite{baier_trust_1986}. In terms of our vision for widespread adoption of AI refactoring, the development community needs a continuous working relationship with AI assistants (the trusters) based on the dependable habits toward the developers, testers, and other roles (the trustees). A well-cited review article by Lee and See presents substantial evidence that trust guides reliance~\cite{lee_trust_2004}, i.e., trust is a meaningful construct for us to study and support. Furthermore, the two authors propose the definition of trust that steers our work: ``the attitude that an agent will help achieve an individual’s goals in a situation characterized by uncertainty and vulnerability.''

\begin{figure}
    \centering
    \includegraphics[width=0.99\columnwidth]{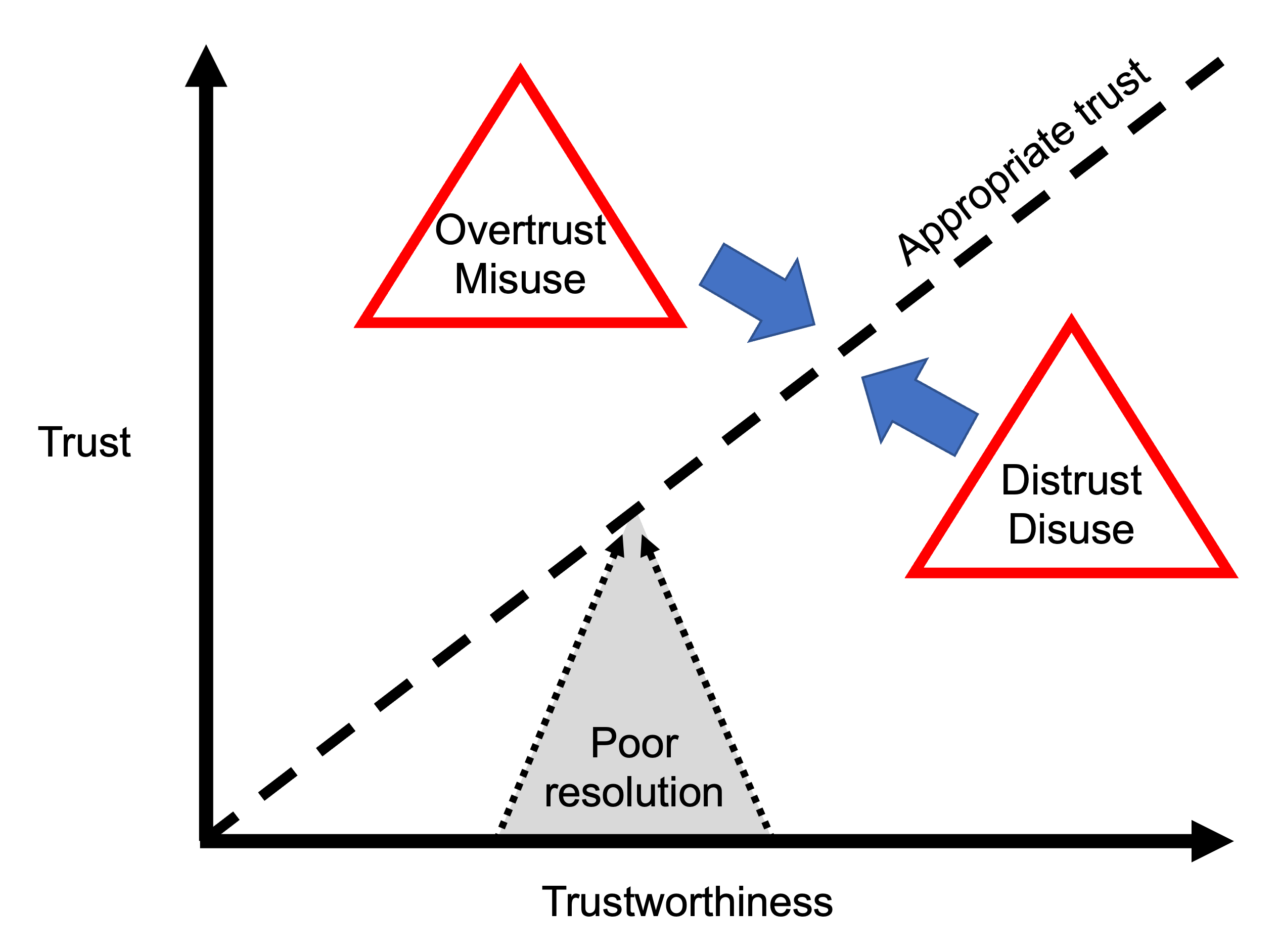}
    \caption{Balancing trust and trustworthiness for AI refactoring.}
    \label{fig:trust-calib}
\end{figure}

Unfortunately, misalignment between human trust and the trustworthiness of automation is common. Fig.~\ref{fig:trust-calib} depicts the phenomenon, based on seminal work by Parasuraman and Riley~\cite{parasuraman_humans_1997}. The terms \textit{misuse} and \textit{disuse} explain failures from flawed partnerships between humans and automation. Misuse refers to the failures that occur when users overtrust automation, i.e., trust $>$ trustworthiness. Disuse denotes failures that occur when people distrust the capabilities of automation, i.e., trust $<$ trustworthiness. The diagonal line represents appropriate trust, a balanced state between trust and trustworthiness, i.e., trust = trustworthiness.

Fig.~\ref{fig:trust-calib} also presents two essential concepts related to obtaining appropriate trust. \textit{Calibration} (blue arrows) is efforts taken to remedy misalignment between trust and trustworthiness. Resolution represents how ``precisely a judgment of trust differentiates levels of automation''~\cite{parasuraman_humans_1997}. The gray area shows an example of poor resolution, as different levels of trustworthiness map to the same trust. In the AI refactoring context, this pertains to differently capable AI assistants being entrusted equally by developers. The poor resolution effectively turns trust into a nonlinear function of trustworthiness. On top of this, human psychology also contributes to non-linearities~\cite{muir_trust_1996}. Examples include the lasting effect of the initial experience and that an AI assistant's worst behaviors disproportionally impact trust -- implications that can persist across several subsequent releases, even with improvements.

\begin{figure}
    \centering
    \includegraphics[width=0.99\columnwidth]{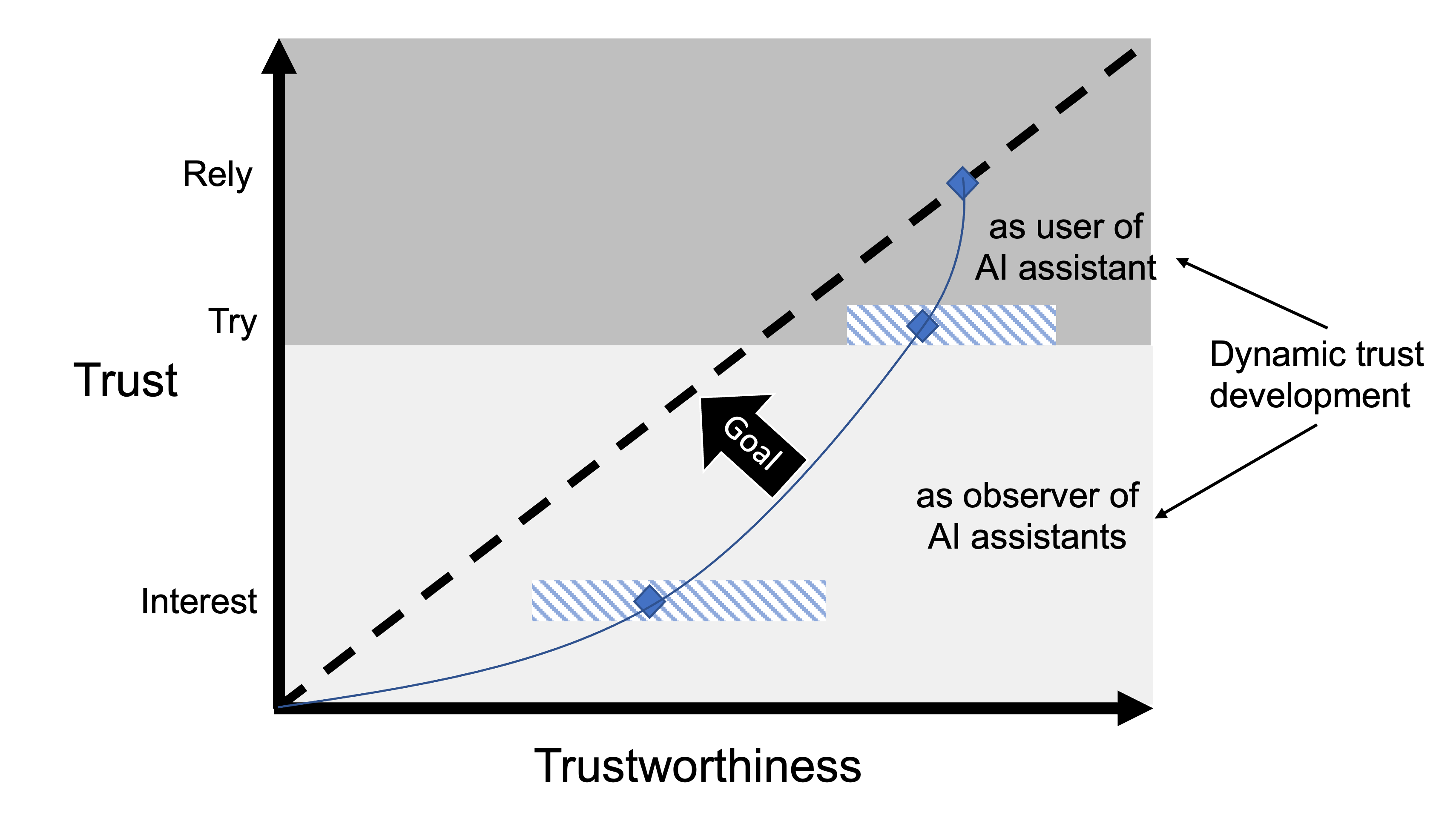}
    \caption{Trust development for a skeptical user of AI refactoring.}
    \label{fig:trust-example}
\end{figure}

Fig.~\ref{fig:trust-example} displays the trust development of a developer that initially distrusts AI assistants. The y-axis indicates three critical breakpoints: I) \textbf{Interest}: recognizing a potential option, II) \textbf{Try}: the willingness to experiment with the tool, and III) \textbf{Rely}: readiness for habitual use. First, the developer's trust development is gradual until reaching I). Trust development is initially gradual until reaching I), after which it accelerates, particularly at II), before eventually reaching III). Striped intervals underscore significant individual variations. The black arrow shows our aim to expedite trust calibration from skepticism to appropriate trust among developers, which will be guided by Hoff and Bashir's conceptual model for dynamic trust development (described in Sec.~\ref{sec:trustdev}).

Parasuraman and Riley also define \textit{specificity} to denote the degree of trust associated with a particular component or aspect of the automation. Specificity is highly relevant to AI refactoring, as the effectiveness of solutions can vary across different code smells. To maximize their usefulness, developers may need support in understanding the specific contexts in which AI assistants excel, rather than treating their performance as uniform. How to best tackle this challenge is another topic for future research.

\section{Trust Development in the IDE} \label{sec:trustdev}
Several researchers have developed models to describe the development of trust in automation. Since trust is a multi-faceted concept, trust development inevitably follows suit. Lee and See discuss how trust evolves as users process information about the capabilities of automation~\cite{lee_trust_2004}, i.e., its trustworthiness. The authors describe how three different types of information processing are involved. Most effective is the \textit{affective process}, i.e., ``emotional responses to violations and confirmations of implicit expectancies.'' The remaining two ways are the \textit{analytical process} (rational evaluation) and the \textit{analogical process} (comparing to opinions of others, including the online developer community~\cite{cheng_it_2024}). Previous work clearly shows that users must both think and feel to develop trust in automation, but feelings dominate -- trust is emotional.

We embark from the body of knowledge on trust development. Primarily, our starting point is a conceptual model of factors that influence trust and reliance in automation developed by Hoff and Bashir~\cite{hoff_trust_2015}. This model, in turn, builds on three different components of trust identified by Marsh and Dibben~\cite{marsh_role_2003}. \textit{Dispositional trust} represents an individual’s enduring tendency to trust automation. This level of trust tends to be stable over time, primarily influenced by culture, age, gender, and personality traits. \textit{Situational trust} reflects how the trust development depends on the situation, i.e., external variability (e.g., type of system, complexity, and task difficulty) and internal variability (e.g., self-confidence and mood). \textit{Learned trust} is split into 1) Initial Learned trust based on pre-existing knowledge before interaction (e.g., experience with similar systems, reputation of the brand, and technology insights) and 2) Dynamic Learned trust that evolves during interaction with the system. 

\begin{figure*}
    \centering
    \includegraphics[width=1.3\columnwidth]{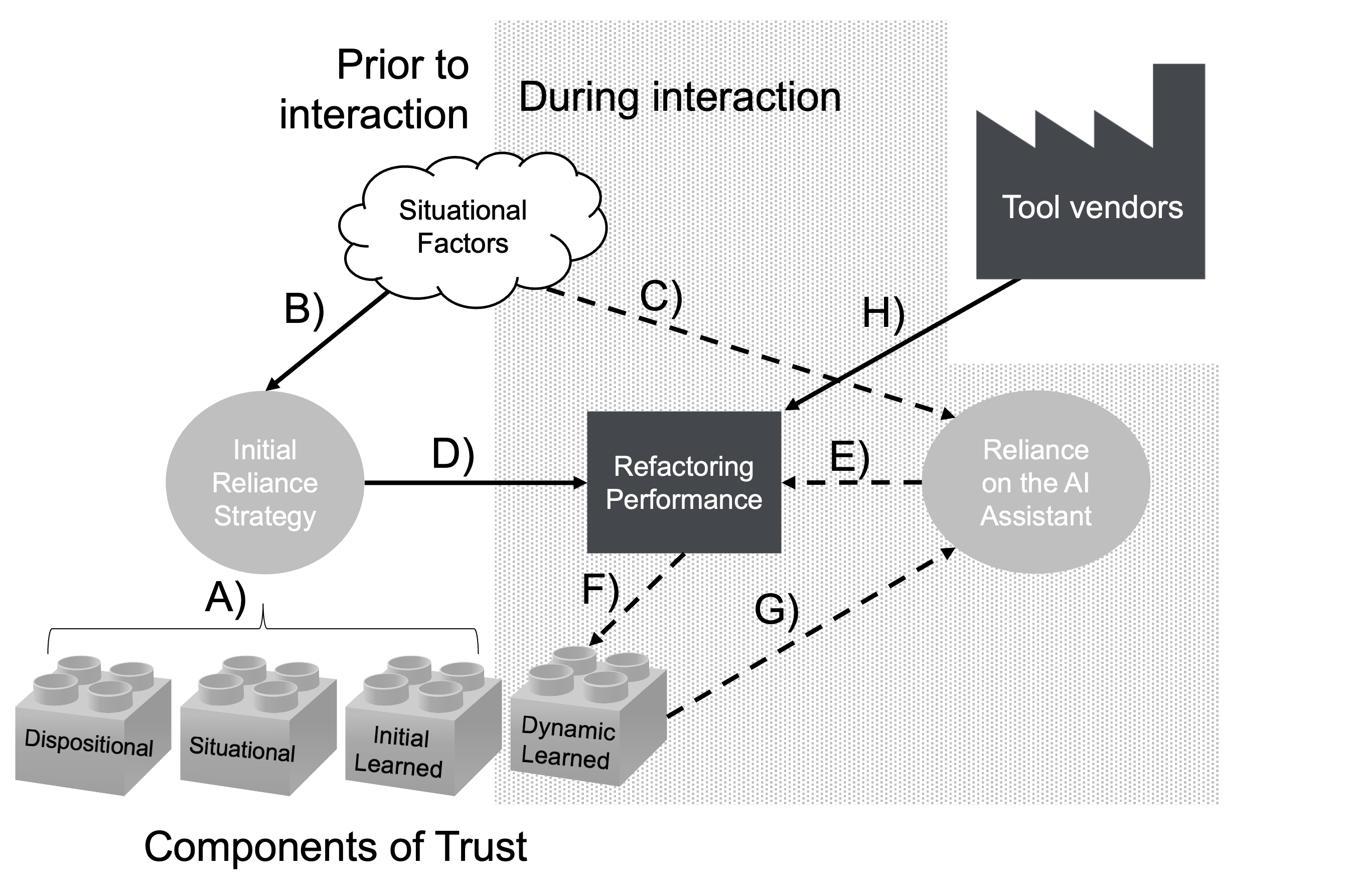}
    \caption{Factors that affect trust in AI refactoring.}
    \label{fig:trust-dev}
\end{figure*}

Fig.~\ref{fig:trust-dev} shows a simplified version of Hoff and Bashir's model~\cite{hoff_trust_2015}. Dashed arrows indicate factors that can change during a single interactive session. The model is primarily organized into factors prior to interaction (the left side) and during interaction (the gray area). A) depicts the three components of trust that develop prior to system interaction, i.e., Dispositional, Situational, and Initial Learned trust. Situational factors that are not directly related to trust can influence both B) the Initial Reliance Strategy prior to interaction and C) Reliance on the AI Assistant during interaction. The Initial Reliance Strategy influences how users interact with the AI Assistant, subsequently D) influencing Refactoring Performance. In the same vein, the user's Reliance on the AI Assistant during interaction E) influences the Refactoring Performance. Users continuously process information related to the Refactoring Performance, which F) affects the Dynamic Learned trust, subsequently G) influencing the Reliance on the AI Assistant. E-F-G constitute a closed loop, illustrating how the user's trust develops as they observe the performance of the automation~\cite{muir_trust_1996}. 

To conclude the section on trust, we paraphrase Lee and See~\cite{lee_trust_2004}: ``appropriate trust and reliance depend on how well the AI refactoring trustworthiness is conveyed to the potential users in the development community.” To support conveyance, Lee and See argue that technology providers should either 1) make the automation simpler or 2) reveal their operation more clearly. As simpler solutions cannot replace the LLMs used in AI refactoring -- opaque AI technology is essential -- there is no alternative to developing trustworthy AI. Researchers must find the appropriate level of detail to explain how AI refactoring work, and customize it based on personal user preferences that dynamically change over time.

\section{Conclusions and the Road Ahead} \label{sec:conc}
Our vision is improved software maintainability thanks to widespread adoption of highly capable AI refactoring. Meeting the vision revolves around scientific foundation around two complementary perspectives of trust. First, \textit{trustworthy} AI refactoring focuses on enhancing the safeguards to increase the refactoring confidence by rejecting low-confidence code editing. Second, \textit{trusted} AI refactoring involves developing the UI to foster developers' trust by customizing the delivery of LLM-based recommendations in the IDE. Previous research on human factors suggests that misalignment between trust and trustworthiness can severely impede industry adoption~\cite{parasuraman_humans_1997} -- we aspire to calibrate these two sides of trust. 

Our research will progress through action research, i.e., ``a disciplined process of inquiry conducted by and for those taking the action''~\cite{staron_action_2020}. Action research is an appropriate method when the main objective is to facilitate change, e.g., improved AI refactoring, and the researcher is an integral part of the change process~\cite{wohlin_guiding_2021}. Implementing research interventions within CodeScene's mature SE intelligence platform offers an opportunity for large-scale evaluations with real users in proprietary projects. Leveraging CodeScene's IDE extensions for VS Code and IntelliJ, we aim to conduct detailed telemetry for A/B testing, enabling continuous enhancement of both the trustworthy safeguards and the trust-cultivating UI, i.e., a user-centered development of our research interventions.

Regarding safeguards, integrating state-of-the-art program analysis~\cite{soderberg_extensible_2013, dura_javadl_2021} and vulnerability analysis~\cite{shen_empirical_2023} are crucial first steps. Furthermore, a range of potential complementary techniques, such as anomaly detection~\cite{henriksson_performance_2021} and mutation analysis~\cite{vercammen_mutation_2024}, will be explored and prioritized as part of our future work. We conclude this position paper with two pivotal research questions that will guide our continued research: 

\begin{itemize}
    \item[RQ\textsubscript{1}] How can we design effective safeguards to enhance the trustworthiness of AI refactoring?
    \item[RQ\textsubscript{2}] In what ways can the UI for AI refactoring be crafted to facilitate trust calibration?
\end{itemize}

\section*{Acknowledgment}
Our thanks go to Christofer Englund and Emma Söderberg for reviewing earlier versions of this work.


\bibliographystyle{IEEEtran}
\bibliography{trust}

\end{document}